\documentclass[aps,prl,twocolumn,superscriptaddress,longbibliography]{revtex4-2}
\pdfoutput=1
\usepackage{bm}
\usepackage{graphicx}
\usepackage{booktabs}
\usepackage{slashed}
\usepackage{hyperref}
\usepackage{amssymb,amsmath,amsfonts}
\usepackage{color}
\usepackage[utf8]{inputenc}
\usepackage[caption=false]{subfig}
\newcommand{\nsat}{n_{\rm sat}}
\newcommand{\ncolor}{N_{\rm color}}

\newcommand{\vphi}{\vec{\phi}}
\newcommand{\vphisq}{\vec{\phi}^{\, 2}}
\newcommand{\tilmom}{\widetilde{m}_\omega}
\newcommand{\meff}{m_{\rm eff}}
\newcommand{\vksq}{\vec{k}^{\, 2}}
\newcommand{\hatom}{\widehat{\omega}_0}
\newcommand{\hateps}{\widehat{\epsilon}}
\newcommand{\qpil}{Q$\pi$L~}
\newcommand{\nfl}{N_{\rm fl}}
\begin{document}

\title{Remarks on nuclear matter: how an $\omega_0$ condensate can spike the speed of sound, and a model of $Z(3)$ baryons
}

\author{Robert D. Pisarski}
\affiliation{Department of Physics, Brookhaven National Laboratory, Upton, NY 11973}

\begin{abstract}

I make two comments about nuclear matter.  First, I consider the effects of 
a coupling between the $O(4)$ chiral field, $\vphi$, and the $\omega_\mu$ meson, $\sim + \, \omega_\mu^2 \, \vphisq$;
for any net baryon density, a condensate for $\omega_0$ is unavoidably generated.
I assume that with increasing density, a decrease of the chiral condensate and 
the effective $\omega_0$ mass gives a stiff equation of state (EoS).
In order to match that onto a soft EoS for quarkyonic matter, I consider
an $O(N)$ field at large $N$, where at nonzero temperature quantum fluctuations 
disorder any putative pion ``condensate'' into a quantum pion liquid (Q$\pi$L) \cite{Pisarski:2020dnx}.
In this paper I show that the \qpil persists at zero temperature.
If valid qualitatively at $N=4$, the $\omega_0$ mass goes up sharply and suppresses the $\omega_0$ condensate.
This could generate a spike in the speed of sound at high density,
which is of relevance to neutron stars.
Second, I propose a toy model of a $Z(3)$ gauge theory with three flavors of fermions,
where $Z(3)$ vortices confine fermions into baryons.  In $1+1$ dimensions this model
can be studied numerically with present techniques, using either classical or quantum computers.

\end{abstract}

\maketitle

To determine the equation of state (EoS) for nuclear matter,
effective models can be used for baryon densities, $n_B$, up to and above
that for nuclear saturation, $\nsat$
\cite{Zeldovich:1962,Serot:1984ey,Walecka:1995mi,Lalazissis:1996rd,Serot:1997xg,Akmal:1998cf,Carriere:2002bx,Danielewicz:2002pu,Epelbaum:2008ga,Machleidt:2011zz}.
Neutron stars probe densities $n_B > \nsat$.
In the past few years, the observation of neutron stars with masses above two solar masses
\cite{Demorest:2010bx,Antoniadis:2013pzd}
and astronomical observations, especially of their mergers in binary systems
\cite{Watts:2016uzu,Ozel:2016oaf,Abbott:2018exr,Miller:2019nzo,Baiotti:2019sew,Annala:2019puf,Oter:2019kig,Drischler:2020hwi,Greif:2020pju,Tews:2018kmu,Greif:2018njt,Forbes:2019xaz,Reed:2019ezm,Drischler:2020fvz,Essick:2020flb,Han:2019bub,Xia:2019xax,Han:2020adu,Kanakis-Pegios:2020jnf,Kojo:2020krb},
have provided a wealth of data.
Many models apply above $\nsat$
\cite{Fukushima:2013rx,Holt:2014hma,Pais:2016xiu,Baym:2017whm,Brown:1991kk,Hatsuda:1991ez,Saito:1996yb,Hayano:2008vn,Gubler:2018ctz,Holt:2017qxy,Rho:2018mni,Ma:2018qkg,Ma:2019ery,Ma:2020hno,Rho:2021zwm,McLerran:2007qj,Andronic:2009gj,Kojo:2009ha,Kojo:2010fe,Kojo:2011cn,Fukushima:2015bda,McLerran:2018hbz,Jeong:2019lhv,Duarte:2020kvi,Duarte:2020xsp,Sen:2020peq,Sen:2020qcd,Zhao:2020dvu,Pisarski:1995xu,Pisarski:1996ne,Detar:1988kn,Jido:2001nt,Zschiesche:2006zj,Gallas:2009qp,Steinheimer:2011ea,Gallas:2011qp,Eser:2015pka,Lakaschus:2018rki,Fattoyev:2017jql,Dexheimer:2018dhb,Malik:2018zcf,Giacosa:2017pos,Suenaga:2019urn,Marczenko:2018jui,Marczenko:2020jma,Cao:2020byn,Minamikawa:2020jfj,Oset:2012ap,Ramos:2013mda,Cabrera:2013zga,Montana:2018bkb,Giacosa:2017mis,Khaidukov:2018vkv,Kojo:2014rca,Baym:2019iky,Song:2019qoh,Fukushima:2020cmk,Leonhardt:2019fua,Otto:2019zjy,Otto:2020hoz,Shahrbaf:2019vtf,Shahrbaf:2020uau,Blacker:2020nlq,Dexheimer:2020rlp,Ecker:2017fyh,Fadafa:2019euu,Kovensky:2020xif,Li:2020dst,Kojo:2020ztt},
but to date, consensus is lacking.

In this paper I make two comments about nuclear matter.  The first is a suggestion as to how $\omega_0$ 
\cite{Zeldovich:1962,Serot:1984ey,Walecka:1995mi,Serot:1997xg}
and pion 
\cite{Overhauser:1960,Migdal:1978az,Kaplan:1986yq,Buballa:2014tba,Azaria:2016,James:2017cpc,Pisarski:2018bct,Pisarski:2020dnx}
condensates can affect the EoS, and generate non-monotonic behavior for the speed of sound
as the density increases well above $\nsat$.
The second is a toy model in which fermion fields, analogous to quarks, are confined into
baryons by a $Z(3)$ gauge field.  

\section{Spiking the speed of sound}

In the past few years astronomical observations of neutron stars
\cite{Demorest:2010bx,Antoniadis:2013pzd,Watts:2016uzu,Ozel:2016oaf,Abbott:2018exr,Miller:2019nzo,Baiotti:2019sew,Annala:2019puf,Oter:2019kig,Drischler:2020hwi,Greif:2020pju,Tews:2018kmu,Greif:2018njt,Forbes:2019xaz,Reed:2019ezm,Drischler:2020fvz,Essick:2020flb,Han:2019bub,Xia:2019xax,Han:2020adu,Kanakis-Pegios:2020jnf,Kojo:2020krb}
have provided significant insight into the nuclear EoS at densities above $\nsat$.  This includes
quantities such as their mass, radius, and tidal deformability.
The EoS is given by the pressure, $p$, as a function of the energy density, $e$.
Analyses with piecewise polytropic EoS are useful \cite{Annala:2019puf,Oter:2019kig,Drischler:2020hwi,Greif:2020pju}.

However, a more sensitive probe of the EoS is given by the speed of sound squared:
$c_s^2 = \partial p/\partial e$.   Free, massless fermions
have $c_s^2 = 1/3$, which is termed soft.  In contrast, several studies of neutron
stars find that it is {\it essential} for the nuclear EoS to have a region in which the EoS is stiff,
where $c_s^2$ significantly larger than $1/3$
\cite{Tews:2018kmu,Greif:2018njt,Forbes:2019xaz,Reed:2019ezm,Drischler:2020fvz,Essick:2020flb,Han:2019bub,Xia:2019xax,Han:2020adu,Kanakis-Pegios:2020jnf,Kojo:2020krb}.

For example, consider the analysis of Drischler {\it et al.} \cite{Drischler:2020fvz}, who extrapolate up from
$\nsat$ using chiral effective field theory.  
To obtain neutron stars with masses above two solar masses, they find that there is a region of density
in which the EoS is stiff: if there is a neutron star of $2.6$ solar masses,
at some $n_B$, $c_s^2 \sim  0.55$.  
To agree with small tidal deformability from GW170817, though, the EoS of nuclear matter must be soft until
$n_B \sim 1.5 - 1.8 \nsat$.

That is, there is a ``spike'' in the speed of sound, with a relatively narrow peak at a density 
significantly above $\nsat$: see, {\it e.g.}, Fig. (1) of Greif {\it et al.} \cite{Greif:2018njt}.

As the density $n_B \rightarrow \infty$, by asymptotic freedom the EoS approaches that of an ideal gas of dense
quarks and gluons, and so is soft.  
In Quantum ChromoDynamics (QCD),
corrections to the quark EoS have been computed in part up to four loop order
\cite{Kurkela:2009gj,Kurkela:2014vha,Kurkela:2016was,Ghisoiu:2016swa,Gorda:2018gpy}.   However, perturbation theory
is only useful down to densities much larger than $\nsat$.  At very high densities, excitations near the Fermi surface
are dominated by color superconductivity \cite{Fukushima:2013rx,Holt:2014hma,Baym:2017whm}.

Going down in density, nuclear matter becomes quarkyonic
\cite{McLerran:2007qj,Andronic:2009gj,Kojo:2009ha,Kojo:2010fe,Kojo:2011cn,Fukushima:2015bda,McLerran:2018hbz,Jeong:2019lhv,Duarte:2020kvi,Duarte:2020xsp,Sen:2020peq,Sen:2020qcd,Zhao:2020dvu}.
The free energy is close to that of QCD perturbation theory, but the excitations near the Fermi surface are confined,
and so baryonic.  A quarkyonic regime is inescapable for a $SU(\ncolor)$ gauge theory as $\ncolor \rightarrow \infty$,
as then quark loops are suppressed by $\sim 1/\ncolor$.
This does not seem to be special to large $\ncolor$, though.
In lattice gauge theory, when $N_{\rm color} \geq 3$ the sign problem prevents classical computers from computing
at zero temperature and nonzero quark density \cite{Banuls:2019rao}.
Two colors, however, is free of the sign problem, and while it 
has unique features - notably, since baryons are bosons there is no Fermi sea - lattice simulations find
a broad quarkyonic region
\cite{Hands:2010gd,Braguta:2016cpw,Bornyakov:2017txe,Boz:2019enj,Philipsen:2019qqm,Astrakhantsev:2020tdl}.
This suggests the same applies to QCD, where $\ncolor = 3$.

I assume that the quarkyonic EoS is soft.
Bedaque and Steiner \cite{Bedaque:2014sqa} have argued, from a variety of examples, the {\it any} quasiparticle
model is soft.  While some authors propose that quarks can give a stiff EoS \cite{Han:2019bub,Han:2020adu,Xia:2019xax},
for simplicity I do not.

To match a nuclear onto a quarkyonic EoS,  McLerran and Reddy take
a quark EoS up to some Fermi momentum $k_{FQ}$, which is then surrounded by a baryonic shell of width $\Delta$,
Fig. (1) of Ref. \cite{McLerran:2018hbz}.  As the baryon density increases, $k_{FQ}$ grows, and $\Delta$ shrinks.
Taking an ideal EoS for both quarks and baryons, 
an appropriate choice of the width $\Delta$ generates a spike in the speed of sound,
Fig. (2) of Ref. \cite{McLerran:2018hbz}.

At densities near $k_{FQ}$, though, {\it neither} equation of state is close to ideal.
The Quantum HadroDynamics (QHD) of Serot and Walecka \cite{Serot:1984ey,Walecka:1995mi,Serot:1997xg} can be
used for baryons, although to be capable of modeling a confined but chirally symmetric phase,
all chiral partners of the nucleons and mesons must be included in a parity-doubled QHD (PdQHD)
\cite{Detar:1988kn,Jido:2001nt,Zschiesche:2006zj,Gallas:2009qp,Steinheimer:2011ea,Gallas:2011qp,Eser:2015pka,Lakaschus:2018rki,Suenaga:2019urn,Marczenko:2018jui,Marczenko:2020jma,Cao:2020byn,Minamikawa:2020jfj}.
The quark EoS can be modeled by coupling quarks and gluons to a linear sigma model for mesons.
Such a PdQHD was considered by Cao and Liao \cite{Cao:2020byn}.

My purpose here is to discuss, in an entirely qualitative manner, of how an $\omega_0$ condensate,
and strong fluctuations in a pion ``condensate''
\cite{Overhauser:1960,Migdal:1978az,Kaplan:1986yq,Buballa:2014tba,Azaria:2016,James:2017cpc,Pisarski:2018bct,Pisarski:2020dnx}
could affect the EoS in PdQHD.  My discussion is admittedly speculative, because given the wealth of experimental data,
it is not easy to describe the EoS of nuclear matter both near $\nsat$ and at $n_B \gg \nsat$.

In QHD, saturation results from a balance between repulsion from the $\omega_\mu$ meson and attraction
from the $\sigma$ meson \cite{Serot:1984ey,Walecka:1995mi,Serot:1997xg}.
That the $\omega_\mu$ meson could generate a stiff EoS was first noted by 
Zel'dovich \cite{Zeldovich:1962}.  Given the coupling of the $\omega_\mu$ to a nucleon
$\psi$ as $\sim g_\omega \, \overline{\psi} \, \omega_\mu \gamma^\mu \, \psi$, then at any nonzero baryon density,
$\langle \overline{\psi} \gamma^0 \psi\rangle = n_B \neq 0$, a condensate for $\omega_0$ is automatically generated
\footnote{As a vector field, an expectation value for $\omega_0$ is generated whenever there is net baryon charge in the system, and is also why
the energy density increases.  This is exactly like QED with net charge, except that here the $\omega_\mu$ meson is massive.
In contrast, for the chiral condensate the energy density is lowered when it has a nonzero expectation value.
I thank L. McLerran for discussions on this point.}:
\begin{equation}
  {\cal L}_\omega^B = - g_\omega n_B \omega_0 + \frac{m_\omega^2 \omega_\mu^2}{2} \Rightarrow
  \langle \omega_0 \rangle = \frac{g_\omega }{m_\omega^2} \, n_B \; .
  \label{omega_lag}
\end{equation}
If only these terms matter, then the EoS is as stiff as possible, with the speed of sound equal to that of light, $c_s^2 = 1$.
Son and Stephanov showed that QCD at nonzero isospin density provides a precise example of this \cite{Son:2000by}.

Of course in QHD, Eq. (\ref{omega_lag}) is not the only term which matters.  Integrating over nucleon loops
at nonzero density, there is an infinite series of terms in $\omega_0$ which are generated at $n_B \neq 0$,
including those $\sim \omega_0^2$, $\sim \omega_0^3$, and so on.  Similarly, the nucleon couples to the
$\sigma$, whose properties also change with $n_B$.  
These effects have been computed to one loop order \cite{Serot:1984ey,Walecka:1995mi,Serot:1997xg}, but
even for strong $g_\omega$, do not dramatically alter the EoS.

The $\omega_\mu$ Lagrangian is
\begin{equation}
  {\cal L}_\omega = \frac{{\cal F}_{\mu \nu}^2 }{4} + \frac{1}{2} \left( \tilmom^2  + \kappa^2 \vphisq\right)\omega_\mu^2 
  \; ;
  \label{omega_lag_mass}
\end{equation}
${\cal F}_{\mu \nu} = \partial_\mu \omega_\nu - \partial_\nu \omega_\mu$ is the field strength for $\omega_\mu$,
$\tilmom$ a mass term, and there is a quartic coupling $\sim \kappa^2$ between
$\omega_\mu$ and $\vphi$, where $\vphi$ is the $O(4)$ chiral field for two light flavors, $\vphi = (\sigma,\vec{\pi})$.
The coupling $\kappa^2$ must be positive to ensure stability for large values of the $\omega_\mu$ and $\vphi$ fields.
(Incidentally, while the term $\sim \tilmom^2$ can be written in 
gauge invariant, unitarity form
\cite{Ruegg:2003ps,Cremmer:1973mg,Hagen:1978jk,Allen:1990gb,Amorim:1995tj,Henneaux:1997mf}, 
that $\sim \kappa^2$ cannot \cite{BF}.)

Although the coupling $\kappa^2$ violates vector meson dominance
\cite{OConnell:1995nse,Rennecke:2015lur,Rennecke:2015eba,Jung:2016yxl,Tripolt:2017zgc,Braun:2017srn,Braun:2018bik,Braun:2019aow,Leonhardt:2019fua,Zhang:2017icm,Otto:2019zjy,Otto:2020hoz}, this is relatively innocuous.  For the $\rho_\mu$ meson, in vacuum a similar
term $\sim \kappa^2 \vec{\rho}_\mu^{\, 2}$ just shifts the $\rho_\mu$ mass, and doesn't alter
its electromagnetic couplings.

In general the complete mass for the $\omega_\mu$ meson is
\begin{equation}
m_\omega^2 = \tilmom^2 + \kappa^2 \, \langle \vphisq \rangle \; .
\label{omega_mass}
\end{equation}
Spontaneous symmetry breaking occurs in the QCD vacuum, $\langle \phi^i \rangle = (\sigma_0,0)$,
where for two flavors, $\sigma_0 = f_\pi$, the pion decay constant.  Thus in vacuum, in mean
field theory the mass squared of
the $\omega_\mu$ meson is $m_\omega^2 = \tilmom^2 + \kappa^2 \sigma_0^2$. 
Large $\tilmom$ favors small $\kappa$, and vice versa.  
For massless pions, at tree level $\tilmom = 0$ when $\kappa^2 = m_\omega/f_\pi \sim 8.4$.

Couplings similar to $\kappa^2$ have appeared before.
In Refs. \cite{Carriere:2002bx} and \cite{Pais:2016xiu}, a chirally asymmetric term $\sim \omega_\mu^2 \sigma^2$ was added to the Lagrangian,
but the generalization to a chirally symmetric term is obvious.  Ref. \cite{Dexheimer:2018dhb} introduced mixing the $\omega_\mu$ and
$\vec{\rho}_\mu$ mesons, $\sim \omega_\mu^2 \vec{\rho}_\mu^{\; 2}$, and note the
possibility of terms $\sim (\omega_\mu^2)^2$, {\it etc.}.  The implications of these terms for neutron stars were computed in Refs.
\cite{Pais:2016xiu,Fattoyev:2017jql,Malik:2018zcf,Dexheimer:2018dhb,Dexheimer:2020rlp}.  
Refs. \cite{Eser:2015pka,Giacosa:2017pos,Lakaschus:2018rki,Suenaga:2019urn} denote $\kappa^2$ as $h_1$, but
neglect it, as $\kappa^2 \sim 1/\ncolor$ is small for a large number of colors.
Ref. \cite{Cao:2020byn} denote $\kappa^2$ as $g_{SV}$; for massive pions they find $\kappa \sim 9.0$, which seems large compared to
the upper bound of $\kappa^2 < 8.4$ for massless pions.

As the density increases, chiral symmetry breaking becomes weaker, and $\sigma_0$ decreases.
Thus Eq. (\ref{omega_mass}) implies that $m_\omega$ decreases with $\sigma_0$, at least
as long as $\langle \vphisq \rangle = \sigma_0^2$.
This is reminiscent of the scaling of Brown and Rho
\cite{Brown:1991kk,Hatsuda:1991ez,Saito:1996yb,Hayano:2008vn,Gubler:2018ctz,Holt:2017qxy,Rho:2018mni,Ma:2018qkg,Ma:2019ery,Ma:2020hno,Rho:2021zwm}.

For the $\vphi$ Lagrangian I take \cite{Pisarski:2020dnx}
\begin{equation}
  {\cal L}_\phi = \frac{( \partial_0 \vphi )^2}{2}  + \frac{( \partial_i^2 \vphi)^2}{2 M^2} 
  + \frac{Z( \partial_i \vphi )^2}{2} 
  + \frac{m_0^2\vphisq}{2} 
  + \frac{\lambda (\vphisq)^2}{4}  \; ,
  \label{lag_phi}
\end{equation}
and work in the chiral limit, so there is no term linear in $\vphi$.  Notice that the $\omega_\mu$ meson does
{\it not} appear
\cite{Pisarski:1995xu,Pisarski:1996ne,Eser:2015pka,Lakaschus:2018rki,Suenaga:2019urn,Marczenko:2018jui,Marczenko:2020jma,Cao:2020byn}.
This is because the $\omega_\mu$ corresponds to the $U(1)_B$ of baryon number, and with $q$ the quark fields,
$\phi \sim \overline{q} q$ is invariant under $U(1)_B$.
Conversely, when $\vec{\rho}_\mu$ and $\vec{a}_1^\mu$ mesons are added, they do appear in Eq. (\ref{lag_phi}), 
since $\vphi$ transforms non-trivially under $SU(2)_L \times SU(2)_R$.  
The $\omega_\mu$ meson does interact with pions through anomalous interactions generated by
the Wess-Zumino-Witten Lagrangian, such as $\omega_\mu \rightarrow 3 \pi$
\cite{Pisarski:1996ne,Eser:2015pka,Lakaschus:2018rki,Suenaga:2019urn,Giacosa:2017pos,Marczenko:2018jui,Marczenko:2020jma,Cao:2020byn,Minamikawa:2020jfj,Oset:2012ap,Ramos:2013mda,Cabrera:2013zga}.
These interactions survive in the chirally symmetric phase, and typically become
$\omega \rightarrow \sigma \pi \pi \pi$, Eq. (15) of \cite{Pisarski:1996ne}.

The anomalous interactions, though, all involve at least three derivatives, which for the $\omega_0$ meson, 
are all spatial derivatives.  
This is why the coupling $\sim \kappa^2$ in Eq. (\ref{omega_lag_mass}) is so important, as
the {\it only} renormalizable, non-derivative coupling which the $\omega_\mu$ has with the chiral field $\vphi$.

This assumes that processes which violate the axial $U(1)_A$ symmetry survive at densities far above $\nsat$,
as indicated by an analysis using a dilute gas of instantons \cite{Pisarski:2019upw}.
If at zero temperature and nonzero baryon density
the axial $U(1)_A$ symmetry remains strongly broken 
by topologically nontrivial configurations even when the
$O(4) = SU(2)_L \times SU(2)_R$ chiral symmetry is restored, 
then as the $\omega_\mu$ meson is a chiral singlet, it need
not become degenerate with its parity partner, which is presumably the
$f_1(1285)$ \cite{Giacosa:2017pos}.  This is unlike mesons which carry flavor, such as the $\rho_\mu$ and $a_1^\mu$,
which are degenerate in a chirally symmetric phase.
Similarly, this is why I can restrict 
the chiral symmetry to be $O(4) = SU(2)_L \times SU(2)_R$ and not $U_A(1) \times SU(2)_L \times SU(2)_R$
\cite{arbNf}.

Of course the $\omega_\mu$ meson interacts directly with nucleons
\cite{Serot:1984ey,Walecka:1995mi,Serot:1997xg,Oset:2012ap,Ramos:2013mda,Cabrera:2013zga}.
The CBELSA/TAPS experiment found that the mass of the $\omega_\mu$
does not shift significantly at nuclear densities
\cite{Trnka:2005ey,Nanova:2010sy}, although its width is over thirty times larger than in vacuum
\cite{Kotulla:2008aa,Metag:2011ji,Thiel:2013cea,Friedrich:2014lba,Metag:2015lza,Metag:2017yuh}.
This does not concur with QHD \cite{Serot:1984ey,Walecka:1995mi,Serot:1997xg} nor
Refs. \cite{Brown:1991kk,Hatsuda:1991ez,Saito:1996yb,Hayano:2008vn,Gubler:2018ctz,Holt:2017qxy,Rho:2018mni,Ma:2018qkg,Ma:2019ery,Ma:2020hno,Rho:2021zwm}, where the $\omega_\mu$ mass decreases by $\nsat$.

This does not exclude changes as the baryon density exceeds $\nsat$.
For the usual analyses of QHD \cite{Serot:1984ey,Walecka:1995mi,Serot:1997xg}, Refs.
\cite{Brown:1991kk,Hatsuda:1991ez,Saito:1996yb,Hayano:2008vn,Gubler:2018ctz,Holt:2017qxy,Rho:2018mni,Ma:2018qkg,Ma:2019ery,Ma:2020hno,Rho:2021zwm},
and PdQHD
\cite{Detar:1988kn,Jido:2001nt,Zschiesche:2006zj,Gallas:2009qp,Steinheimer:2011ea,Gallas:2011qp,Eser:2015pka,Lakaschus:2018rki,Suenaga:2019urn,Giacosa:2017pos,Marczenko:2018jui,Marczenko:2020jma,Cao:2020byn,Minamikawa:2020jfj},
the $\sigma$ and $\omega_0$ masses both decrease, as the balance between $\sigma$ attraction and $\omega_0$ repulsion
gives a soft EoS.

My principal assumption is that for some $n_B > n_1 > \nsat$, that one enters a region dominated by the
$\omega_0$ condensate.  Notably, if $Z$ decreases with increasing $n_B$, 
the effective mass squared of the $\sigma$ increases as $\sim 1/Z$, while if $\kappa \neq 0$,
the $\omega_0$ becomes light as the chiral symmetry is restored.
By  Eq. (\ref{omega_lag}), ${\cal L}_\omega^B = - g_\omega^2 n_B^2/(2 m_\omega^2)$, and a heavy $\sigma$, with a light $\omega_0$,
could generate a stiff EoS for $n_B > n_1$.

{\it Assuming} that a light $\omega_0$ gives a stiff EoS,
then {\it how} can the $\omega_0$ condensate evaporate to match onto a soft quarkyonic EoS?
Presumably the couplings of the $\omega_\mu$ with nucleons behave smoothly with density.
That leaves the couplings of the $\omega_\mu$ to the chiral field $\vphi$, but as demonstrated above, these are limited.
This question does assume that a hadronic phase matches onto quarkyonic matter.  
It is possible to simply paste a stiff hadronic EoS onto a soft quark EoS through what is presumably
a strongly first order transition.  This is not consistent, however, with the analyses for either $\ncolor \rightarrow \infty$
\cite{McLerran:2007qj,Andronic:2009gj,Kojo:2009ha,Kojo:2010fe,Kojo:2011cn,Fukushima:2015bda,McLerran:2018hbz,Jeong:2019lhv,Duarte:2020kvi,Duarte:2020xsp,Sen:2020peq,Sen:2020qcd,Zhao:2020dvu}
or lattice results for $\ncolor = 2$
\cite{Hands:2010gd,Braguta:2016cpw,Bornyakov:2017txe,Boz:2019enj,Philipsen:2019qqm,Astrakhantsev:2020tdl},
which indicate a quarkyonic regime.
Nor why the nuclear EoS appears to be soft near $\nsat$, and only stiff when $n_B \sim 1.5 - 1.8 \nsat$ \cite{Drischler:2020fvz}.

I stress that reducing the contribution of the $\omega_0$ condensate at large chemical potential, $\mu$, and low temperature, $\mu \gg T $, has
{\it no} analogy to the more familiar case, at nonzero temperature and low density.  When $T \gg \mu$,
it is easy matching the EoS of hadronic matter, with a relatively few degrees of freedom,
onto a quark-gluon plasma, with many.
This is precise in the limit of a large number of colors, $\ncolor \rightarrow \infty$, where the pressure in the hadronic phsae 
is $\sim \ncolor^0$, versus $\sim \ncolor^2$ in the deconfined phase.  Similarly, the contribution of the chiral condensate is only
$\sim \ncolor^1$, and decreases as $T$ increases.  In contrast, at $\mu \gg T$, the pressure is always $\sim N_c^1$, in both
the hadronic and quark-gluon phases.

At nonzero density, the appearance of a condensate for $\omega_0$ is special to the $\omega_\mu$ meson:
there is {\it no} other hadron which couples directly to the net baryon density.
This assumes that the only net charge is for baryon number.  When there is a net isospin charge,
a condensate for the $\vec{\rho}_\mu$ meson is generated, $\sim \rho_0^3$.
In this case, terms such as $\vec{\phi}^2 \vec{\rho}_\mu^{\; 2}$, amongst others
\cite{Eser:2015pka,Giacosa:2017pos,Lakaschus:2018rki,Suenaga:2019urn},
need to be included;  further, couplings between the $\omega_\mu$ and $\vec{\rho}_\mu$ mesons, 
$\sim \omega_\mu^2 \vec{\rho}_\mu^{\; 2}$,  must be added \cite{Dexheimer:2018dhb}.  

It is then {\it very} difficult to fit the EoS of an $\omega_0$ condensate
onto that of cold quarks: either the coupling of the $\omega_0$ becomes small, or the mass
of the $\omega_0$ becomes large.   Since the coupling of the $\omega_0$ is strong in vacuum,
the former is most implausible.  Thus the mass of the $\omega_0$ must increase, although this does {\it not} occur in mean field theory  \cite{dexheimer}.
I now argue that the mass of the $\omega_0$ increases sharply due to large quantum fluctuations.

Returning to the Lagrangian in Eq. (\ref{lag_phi}), it 
is standard except for the term quartic in the spatial derivatives, $\sim 1/M^2$ \cite{higher}.
Causality implies that only terms with two time derivatives enter.  With
the term $\sim 1/M^2$ to ensure stability, it is possible to
allow the coefficient of the term with two spatial derivatives, $Z$, to be negative.

While in vacuum $Z=1$ by Lorentz covariance, this is not true in a medium.  If $Z$
is negative, classically a condensate is generated:
\begin{equation}
  \vphi = \sigma_0 (\cos(k_c z),\sin(k_c z),0,0) \; ; \; k_c^2 = \frac{-Z M^2}{2} \; ,
  \label{single_mode}
\end{equation}
which  is a pion condensate in the $z$ direction
\cite{Overhauser:1960,Migdal:1978az,Kaplan:1986yq,Bringoltz:2009ym,Buballa:2014tba,Azaria:2016,James:2017cpc,Pisarski:2018bct,Pisarski:2020dnx}.
In $1+1$ dimensions, such chiral spiral condensates are {\it ubiquitous}
at low temperature and nonzero density
\cite{Buballa:2014tba,Azaria:2016,James:2017cpc}, although in general the solutions are more involved.
Given these examples, it is natural to assume that in QCD, at low temperature $Z < 0$
for some range in density above $\nsat$.

Most discussions of a pion condensate use a nonlinear Lagrangian, in which the $\sigma$ meson does not explicitly appear.
The advantage of using a linear Lagrangian is that it is much easier studying how the symmetric phase is approached.
Following Ref. \cite{Pisarski:2020dnx} I
generalize from $O(4)$ to $O(N)$, where the solution is direct as $N \rightarrow \infty$  \cite{arbNf,NandNf}.

The solution at large $N$ is standard, and proceeds by introducing the a field $\xi = \vphisq$, and a constraint
field, $\epsilon$,
${\cal L}_{\rm cons} = i \epsilon(\xi - \vphisq)/2$.
I only seek the solution for the symmetric phase, although the solution in the broken phase can also be determined
\cite{Pisarski:2020dnx}.
Using this constraint, the $\vphi$ and $\xi$ fields are integrated out to give the effective action
\begin{equation}
  {\cal S}_{\rm eff} = \frac{N}{2} {\rm tr} \log \Delta^{-1} 
+ \int d^4 x ( \frac{ \epsilon^2}{4 \lambda} + \frac{\tilmom^2 \omega_\mu^2}{2} - g_\omega \omega_0 \rho_B ) ;
\end{equation}
$\Delta^{-1}$ is the inverse propagator for the $\vphi$ field, which 
in momentum space is $\Delta^{-1}(\omega,\vec{k}) = \omega^2 + E(k)^2$, where
\begin{equation}
  E(k)^2 = \frac{(\vksq)^2}{M^2} + Z \vksq + \meff^2 \; ,
 \end{equation}
I expand about a stationary point in $\epsilon$ and $\omega_0$, $\epsilon = i \hateps + \epsilon_q$
and $\omega_0 = \hatom + \omega_0^q$, where $\epsilon_q$ and $\omega_0^q$ are quantum fluctuations.
The effective mass $\meff^2 = m_0^2 + \hateps + \kappa^2 \hatom^2$.  
To have a well defined limit for large $N$, as $N \rightarrow \infty$ all terms in the action should
scale as $\sim N$, so I take $\lambda, \kappa^2 \sim 1/N$,
$g_\omega \rho_B , \hatom \sim \sqrt{N}$, and
$\tilmom^2, M^2 , Z, m_0^2, \hateps ,\meff^2\sim N^0$.  Remember that $N$ is just a fictitious parameter,
and is not related to the number of colors or flavors.

Requiring that the effective action is a stationary point in $\epsilon_q$ and $\omega_0^q$
fixes $\hateps$ and $\hatom$,
\begin{equation}
  \hateps = \lambda N {\rm tr} \Delta \;\; ; \;\;
  \hatom = \frac{g_\omega \rho_B}{\tilmom^2 + \kappa^2 N {\rm tr} \Delta} \; .
  \label{stat_pt}
  \end{equation}

  The solution for general values of the parameters is involved, so to make a qualitative point
  I only consider the limit of $Z \rightarrow - \infty$, where classically 
the condensate of Eq. (\ref{single_mode}) dominates.
Instead, in perturbation theory one finds that would be Goldstone bosons have a double pole at {\it non-}zero momentum, about
$k_c$ \cite{Pisarski:2020dnx}.  Such a double pole generates a logarithmic infrared divergence at zero temperature,
and a power law divergence at nonzero temperature.

The solution at large $N$ shows how these infrared divergences are avoided.
As $Z \rightarrow -\infty$, at $N = \infty$ take $\meff \approx -Z M/2 + \delta\meff$.
About $k \approx k_c$, 
\begin{equation}
  E(k \approx k_c)^2 \approx  \frac{1}{M^2} ( (k^2 - k_c^2)^2 - Z M^3 \delta \meff + \ldots) \; .
  \label{disp_kc}
\end{equation}
The loop integral is dominated by $k \approx k_c + \delta k$, and to leading logarithmic order becomes
\begin{eqnarray}
  {\rm tr} \Delta &\approx& \frac{\sqrt{-Z} M^2}{2^{5/2} \pi^2} \int \frac{d \delta k}{\sqrt{(\delta k)^2 + M \delta \meff/2}}
                            \nonumber \\
    &\approx& \sqrt{-Z} \frac{M^2}{2^{7/2} \pi} \log\left(\frac{\# \sqrt{-Z} M}{\delta \meff}\right) \; . \\
\end{eqnarray}
Solving Eq. (\ref{stat_pt}) for $\hateps$, as $Z \rightarrow -\infty$,
\begin{equation}
  \delta \meff \approx \# \sqrt{-Z} M \exp\left( - \; \frac{2^{3/2} \pi^2}{\lambda N} (-Z)^{3/2} \right)
  \;  ,
\end{equation}
where $\#$ is a positive, nonzero number.

It is worth contrasting this solution with that at nonzero temperature \cite{Pisarski:2020dnx}.  Then
the integral over $\omega$ is a discrete sum, and the zero energy mode is the most important.  It generates
a power law divergence, with the solution $\delta \meff \approx 1/Z^4$,
Eq. (58) of Ref. \cite{Pisarski:2020dnx}.  The statement in Ref. \cite{Pisarski:2020dnx}
that $\delta \meff$ vanishes at zero temperature is incorrect: it is just that $\delta \meff$ is suppressed
exponentially in $1/\sqrt{-Z}$, instead of by a power.  I refer to this disorder as a quantum pion liquid, Q$\pi$L \cite{name}.

I have neglected the equation for $\hatom$ in Eq. (\ref{stat_pt}).  
While $\delta \meff^2$ is very different at zero and nonzero temperature, though, 
what matters there is the value of the loop integral.
Since $\hateps \approx \meff^2 \sim Z^2 M^2/4$, by Eq. (\ref{stat_pt})
\begin{equation}
  {\rm tr} \Delta \approx Z^2 \; \frac{M^2}{4 \lambda N} \; , \; Z \rightarrow -\infty \; ,
\end{equation}
As $\langle \vphisq \rangle = N {\rm tr} \Delta$,
by Eq. (\ref{omega_mass}) the $\omega_0$ mass increases sharply,
\begin{equation}
  m_\omega^2 = \tilmom^2 + Z^2 \; \frac{\kappa^2 }{4 \lambda} \; M^2 \; ,
\label{omega_mass_qsl}
\end{equation}
and by Eq. (\ref{stat_pt}) suppresses the $\omega_0$ condensate, $\hatom \sim 1/Z^2$.  Note that
the presence of the coupling $\sim \kappa^2$ is {\it essential} for this to occur.

When $Z$ is negative, classically one expects a pion condensate to form,
but the solution at large $N$ shows that instead a quantum pion liquid (Q$\pi$L) forms .
While this is rigorous at large $N$, as it arises from the double pole at $k_c \neq 0$ for
the would be Goldstone modes, it is very likely that there is a \qpil for {\it all} $N > 2$
\cite{Pisarski:2020dnx}.  I also assume that the quantum fluctuations are sufficiently strong
so that a \qpil forms for massive pions.

My suggestion is thus the following.  For $n_B > n_1 > \nsat$, the theory enters a phase dominated by the $\omega_0$
condensate, which stiffens the EoS.  When $n_B > n_2$, it is approximately described by a Q$\pi$L:
both the $\sigma$ and $\omega_0$ are heavy, which suppresses $\hatom$.  In total, the enhancement and then
suppression of the $\omega_0$ condensate generates a spike in the speed of sound.

Clearly a detailed analysis is required to determine the dependence of the various parameters with density,
or more properly for thermodynamics, with the baryon chemical potential, $\mu_B$.
This includes the $\mu_B$ dependence of the wave function renormalization constant $Z$,
the mass parameter $M$ (which is of some hadronic scale), $m_0$, $\lambda$, and so forth.

The most direct approach is to use PdQHD, with a self-consistent
one loop approximation for the nucleons, the chiral fields $\vphi$, and the $\omega_0$.
While involved, I comment that it is far 
simpler to look for a \qpil - which is just a non-monotonic dispersion relation, Eq.
(\ref{disp_kc}) - than for a pion condensate, which is not spatially homogeneous \cite{Buballa:2014tba}.

As quantum computers are (very) far from computing the properties of cold, dense QCD \cite{Banuls:2019rao},
to proceed from first principles requires the functional renormalization group (FRG)
\cite{Rennecke:2015lur,Rennecke:2015eba,Jung:2016yxl,Tripolt:2017zgc,Braun:2017srn,Braun:2018bik,Braun:2019aow,Leonhardt:2019fua,Zhang:2017icm,Otto:2019zjy,Otto:2020hoz,Braun:2017srn,Braun:2018bik,Braun:2019aow,Mitter:2014wpa,Cyrol:2016tym,Cyrol:2017ewj,Fu:2019hdw}.
Ref. \cite{Leonhardt:2019fua} use a chiral effective model up to $\sim 2 \nsat$,
matching onto QCD perturbation theory with a Fierz complete FRG
\cite{Braun:2017srn,Braun:2018bik,Braun:2019aow}
at intermediate $n_B$.  They find evidence for a spike in the speed of sound 
at $\sim 10 \nsat$ \cite{Leonhardt:2019fua}, which is much higher than Ref. \cite{Drischler:2020fvz}.
The ultimate goal is to use the parameters determined by the FRG in vacuum 
\cite{Mitter:2014wpa,Cyrol:2016tym,Cyrol:2017ewj} to compute the EoS for nuclear matter.  
Fu, Pawlowski, and Rennecke \cite{Fu:2019hdw} find that $Z < 0$ at rather high $T$ and $\mu_B \neq 0$,
Fig. (21) of \cite{Fu:2019hdw}.  A complete FRG analysis should certainly see a quantum pion liquid, if it exists.

The pion is not an exact Goldstone boson, but I assume it is
so light that the \qpil phase wins over a pion condensate.
The same may not be true for strange quarks \cite{Tolos:2020aln}.
When at some $n_B$ the Fermi sea spills over to form one of strange quarks, if the pion $Z$ is negative, by
$SU(3)$ flavor symmetry that for kaons will be as well.
As the strange quark is much heavier than up and down quarks,
instead of a quantum kaon liquid, a kaon condensate {\it might} form \cite{Kaplan:1986yq}.
This would be a crystal of real kinks, where
$\langle \overline{s} s \rangle$ oscillates about a constant, nonzero value
Bringoltz \cite{Bringoltz:2009ym} showed that this happens for 
the 't Hooft model in $1+1$ dimensions \cite{tHooft:1974pnl}.

Admittedly my analysis is merely a sketch of how a spike in the speed of sound
might arise in nuclear matter.  It appears inescapable, though, that the interaction of the $\omega_0$
and the chiral fields plays an essential role.  

\section{A model of $Z(3)$ baryons}

Some properties of nuclear matter, such as those discussed above, are surely special to QCD.
It would be useful, however, to have the simplest possible model
which exhibits the confinement of some type of ``quarks'' into baryons.
A $SU(\ncolor)$ gauge theory in $1+1$ dimensions \cite{tHooft:1974pnl}
has baryons \cite{Steinhardt:1980ry,Affleck:1985wa,Abdalla:1995dm,Bringoltz:2009ym}, but as $\ncolor \rightarrow \infty$,
there are $\sim \ncolor^2$
degrees of freedom.  There are also models in $1+1$ dimensions which are soluble about the conformal limit 
\cite{Azaria:2016,James:2017cpc}, but these do not generalize to higher dimensions.

An understanding of
confinement from $Z(\ncolor)$ vortices in a $SU(\ncolor)$ gauge theory was proposed by 't Hooft
\cite{tHooft:1977nqb,tHooft:1979rtg}; for recent work, see
\cite{Greensite:2011zz,Greensite:2016pfc,Biddle:2019gke} and references therein.
I suggest discarding the non-Abelian degrees of freedom in $SU(\ncolor)$ to retain just those of $Z(\ncolor)$.
A $Z(3)$ gauge theory is constructed following Krauss, Wilczek, and Preskill
\cite{Krauss:1988zc,Preskill:1990bm,deWildPropitius:1995hk}:
\begin{eqnarray}
  {\cal L}_{Z(3)} &=& \frac{F^2_{\mu \nu}}{4} + |D_\mu^\chi \chi|^2
                      + m_\chi^2 |\chi|^2 + \lambda_\chi (|\chi|^2)^2 \nonumber\\
                  &+&  \sum_{i=1}^3 \overline{q}_i (D_\mu + m_q) q_i \; ,\\
  \nonumber
  \label{z3_gauge}
\end{eqnarray}
where $F_{\mu \nu} = \partial_\mu A_\nu - \partial_\nu A_\mu$ is the field strength for an Abelian gauge field;
$\chi$ is a complex valued scalar, and there are three degenerate types, or ``flavors'' of fermions, $q_i$, with
equal mass $m_q$.
The $q_i$ have unit charge, $D_\mu = \partial_\mu - i e A_\mu$, but I choose the scalar to have charge three,
$D^\chi_\mu = \partial_\mu - 3 i e A_\mu$.

I consider the case of $1+1$ dimensions first, and assume that the fermions are heavy.
(Light fermions, $m_q \ll e$, may undergo spontaneous symmetry
breaking, which because of the lack of Goldstone bosons in $1+1$ dimensions,
complicates the analysis
\cite{Steinhardt:1980ry,Affleck:1985wa,Abdalla:1995dm,Azaria:2016,James:2017cpc},
and is really secondary to my desire to construct a theory for nuclear matter.)
If $m_\chi^2 < 0$, spontaneous symmetry breaking occurs, and the photon becomes massive.  For large distances,
$> 1/(3 e |m_\chi|)$, naively one expects that there is no interaction from the photons, and the fermions
propagate freely.  
Besides perturbative fluctuations, there are also vortices, which in two (Euclidean) dimensions are like
pseudoparticles, localized at a given point.  The vacuum is a superposition of
vortices, where each vortex has an action
$S_v \sim (m_\chi)^4/\lambda_\chi$.  If $\chi$ had unit charge, the propagation of fermions is affected only when
they are near a vortex, and the vortices are relatively inconsequential.

When $\chi$ has charge three, however, a vortex can carry a $Z(3)$ charge, which greatly affects the propagation
of the fermion.  If a fermion of unit charge encircles a single vortex, it picks up an Aharonov-Bohm phase of
$\exp(\pm 2 \pi i/3)$.  With a vacuum composed of an infinite number of vortices, these phases confine
\footnote{My language is common but imprecise: since the fermions carry $Z(3)$ charge, any flux string can always break by the production of fermion
anti-fermion pairs.   By confinement I mean that there is a mass gap for all particles in the spectrum, which by necessity are gauge singlets.},
the fermions
entirely through these random phases, exactly analogous to how $Z(3)$ vortices in a $SU(3)$ gauge theory
confine \cite{Greensite:2016pfc}.

While a state such as $q_1^3$ is neutral under $Z(3)$, this vanishes, as $q_1$ is a fermion
field which anti-commutes with itself.  This is different from QCD, where the anti-symmetric tensor in color
space can be used to form a baryon with one flavor, $\sim \epsilon^{a b c}q_1^a q_1^b q_1^c$.  Consequently,
in a $Z(3)$ model to obtain (simple) baryons it is necessary to take three flavors, so the baryon
$\sim q_1 q_2 q_3$, is neutral under $Z(3)$.  The mesons form an octet in flavor, which is (presumably) lighter than
the singlet meson (plus higher excitations, of course).

In weak coupling the action of a single vortex is small, vortices are dilute, and confinement occurs over large
distances, $\sim \exp(-S_v)$.   The fermions interact over distance $\sim 1/m_\chi$, but at long distances, only
interact through the $Z(3)$ phases generated by the vortex ensemble in vacuum.  These $Z(3)$ baryons are 
weakly bound over large distances, so that in any scattering experiment, it would
be obvious that they have composite substructure.  This is in contrast to QCD, where baryons have weak attraction
at large distances, but a strong repulsive core at short distances.

That is, in QCD it is hard prying the quarks out of a baryon.  This would occur if the density of vortices is
large.  In the effective model above, this requires strong coupling, which cannot be studied analytically.
However, this limit can be studied on the lattice, and just produces a $Z(3)$ gauge theory
\cite{Horn:1979fy} coupled to three
flavors of degenerate fermions.

In $1+1$ dimensions, as for the $U(1)$ gauge theory \cite{Coleman:1975pw,Coleman:1976uz}, the $Z(N)$
gauge theory confines.  On a lattice, classical computers have been used to study the properties in vacuum of
$Z(2)$ \cite{Zohar:2016wmo,Ercolessi:2017jbi,Borla:2019chl,Frank:2019jzv,Magnifico:2019ulp}
and $Z(3)$ \cite{Ercolessi:2017jbi,Magnifico:2019ulp} gauge theories with a single flavor.
The behavior of a $U(1)$ theory with two flavors was computed at non-zero density in Ref. \cite{Banuls:2016gid}.
Thus classical computers can be used to compute the properties of
a $Z(3)$ gauge theory with three degenerate flavors at nonzero density.  This can then provide a benchmark
to compare against computing the free energy at nonzero density using quantum computers.  The great advantage of
a $Z(3)$ gauge group is that only two qubits are needed to describe a group element, as opposed to many
more for any continuous gauge group.  

In $2+1$ dimensions the vortices sweep our lines in space-time, and cylinders in $3+1$ dimensions.  Assuming that
$Z(3)$ vortices confine in QCD, these models should exhibit confinement as well.  It would be interesting analyzing
the behavior of $Z(3)$ nuclear matter at strong coupling as a counterpoint to that in QCD.

\acknowledgments
I thank S. Carignano, L. Classen, V. Dexheimer, T. Izubuchi, Y. Kikuchi, Y.-L. Ma, L. McLerran, F. Rennecke,
M. Rho, D. Rischke, F. Tartaglia, A. Tsvelik, and A. Tomiya
for discusions; also Brian Serot, albeit some time ago, about the ineluctable virtues of cockatiels and QHD;
and S. Reddy, for discussions on the role of an $\omega_0$ condensate.
This research was supported by the U.S. Department of Energy 
under contract DE-SC0012704; the work in
Sec. I, by B.N.L. under the Lab Directed Research and Development program 18-036; the work in
Sec. II, by the U.S. Department of Energy, Office of Science
National Quantum Information Science Research Centers under the award for the ``Co-design Center for Quantum Advantage''.
After this work was completed, the nature of the Q$\pi$L regime was developed further with
A. Tsvelik \cite{Tsvelik:2021ccp};
the signatures of a Q$\pi$L regime in heavy ion collisions were analyzed with F. Rennecke \cite{Pisarski:2021qof}.

%\bibliography{lifshitz}

\begin{thebibliography}{184}
\bibitem [{arb()}]{arbNf}%
  \BibitemOpen
  \href@noop {} {}\bibinfo {note} {For an arbitrary number of flavors, $\nfl$,
  under a chiral rotation the chiral field $\Phi$ transforms as $\Phi
  \rightarrow {\rm e}^{i \theta_A} \, {\rm e}^{i \alpha_L} \Phi {\rm e}^{- i
  \alpha_R}$, where $\theta_A$, $\alpha_L$ and $\alpha_R$ are elements of the
  Lie algebra for $U_A(1) \times SU(\nfl)_L \times SU(\nfl)_R$, respectively.
  For arbitrary $\nfl$ the coupling $\sim \kappa^2$ in Eq.
  (\ref{omega_lag_mass}) generalizes trivially to $\omega_\mu^2 \, {\rm tr}
  \Phi^\dagger \Phi$.}\BibitemShut {Stop}%
\bibitem [{Nan()}]{NandNf}%
  \BibitemOpen
  \href@noop {} {}\bibinfo {note} {For arbitary $\nfl$, $\Phi$ is a complex,
  $\nfl \times \nfl$ matrix. While in general matrix models are not easily
  solved at large $\nfl$, for the ansatz where $SU(\nfl)_L \times SU(\nfl)_R$
  breaks to $SU(\nfl)_V$, which is $\Phi = \sigma_0 {\bf 1}$, the solution
  follows directly from the vector model, with $N \rightarrow 2
  \nfl^2$.}\BibitemShut {Stop}%
\end{thebibliography}

\end{document}